\documentclass[11pt,draftcls,onecolumn,journal]{IEEEtran}
 \usepackage{graphicx}
\usepackage{epstopdf}

\usepackage{textcomp}
\usepackage{stfloats}
\usepackage{url}
\usepackage{verbatim}
\usepackage{graphicx}
\usepackage{cite}
\usepackage{hyperref}
\usepackage{booktabs}
\usepackage{amsmath,epsfig,amssymb,verbatim,amsopn,cite,multirow}
\usepackage{amsthm}
\usepackage{balance}
\usepackage{multirow}
\usepackage[usenames,dvipsnames]{color}
\usepackage[all]{xy}  
\usepackage{url}
\usepackage{amsfonts}
\usepackage{amssymb}
\usepackage{epsfig}
\usepackage{epstopdf}
\usepackage{bm}
\usepackage{balance}
\usepackage{array}
\usepackage{graphicx}
\usepackage{subcaption}
\usepackage{footnote}
\usepackage{algorithm,algorithmic}
\usepackage{amssymb}

\usepackage{makecell,multirow}
\usepackage{rotating}
\usepackage{tcolorbox}

  {\proof}{\proofend}

\newcounter{mytempeqcounter}

\newcommand{\Ex}{\mathbb{E}}

\newcommand{\qg}{{\bf g}}
\newcommand{\qh}{{\bf h}}

\newcommand{\qn}{{\bf n}}

\newcommand{\qq}{{\bf q}}

\newcommand{\qw}{{\bf w}}
\newcommand{\qx}{{\bf x}}
\newcommand{\qy}{{\bf y}}

\newcommand{\qG}{{\bf G}}

\newcommand{\Ntx}{M}

\newcommand{\Ins}{\mathtt{ins}}
\newcommand{\Stat}{\mathtt{stat}}

\newcommand{\rod}{\rho_\mathrm{d}}

\newcommand{\ghat}{\hat{\qg}}

\newcommand{\SINRk}{\mathrm{SINR}_{k}}

\newcommand{\sti}{\mathrm{i}}

\begin{document}
%
\title{Massive MIMO: Instantaneous versus Statistical CSI-Based Power Allocation}
%
\author{Zahra~Mobini,~\IEEEmembership{Member,~IEEE,}
Hien~Quoc~Ngo,~\IEEEmembership{Fellow,~IEEE}

\thanks{Z. Mobini and H. Q. Ngo  (corresponding author)  are with the Centre for Wireless Innovation (CWI), Queen's University Belfast, BT3 9DT Belfast, U.K. (email:\{zahra.mobini, hien.ngo\}@qub.ac.uk). H. Q. Ngo is also with the Department of Electronic Engineering, Kyung Hee University, Yongin-si, Gyeonggi-do 17104, South Korea.}
}
%
\markboth{IEEE Signal Processing Magazine,~Vol.~XX, No.~XX, June~2024}%
{Comminiello \MakeLowercase{\textit{et al.}}: Author Guidelines for Columns \& Forum Articles of IEEE SPM}


\maketitle
%
%
%
%
%
\section*{}
\label{sec:abstract}
Massive multiple-input multiple-output (mMIMO) has emerged as a key enabler for the fifth generation (5G) and beyond wireless communications, offering the potential to meet the ever-growing demands of an increasingly connected world. Unlike traditional MIMO systems that employ a limited number of antennas at the base stations (BSs), mMIMO utilizes a significantly larger number of antennas, often numbering in the hundreds or even thousands. This massive antenna deployment enables simultaneously serving many tens of users without any need to additional frequency resources. Overall, by leveraging spatial multiplexing and precoding/combining techniques in conjunction with efficient power allocation, mMIMO can mitigate interference, enhance signal quality, and hence,  achieve remarkable gains in both spectral and energy efficiency.
More specifically, the optimization of power allocation plays a crucial role in minimizing interference within mMIMO systems, offering two principal implementation avenues: instantaneous channel state information (CSI)- and statistical CSI-based schemes.
 Instantaneous CSI-based power control schemes exhibit notable performance gains, particularly in scenarios characterized by significant variations in channel quality across individual subcarriers. These variations arise due to differing channel conditions, such as fading that affect each subcarrier independently. By adapting power allocation based on the instantaneous  CSI, the system can more effectively optimize transmission parameters, improving overall system performance.
 The deployment of instantaneous CSI-based power control schemes necessitates computationally intensive signal processing operations, requiring substantial resources to handle real-time CSI updates and the associated overhead.
Conversely, statistical CSI-based schemes enable  efficient implementation of advanced power allocation algorithms within large-scale mMIMO systems, where the algorithms are updated much less frequently. Nevertheless, these schemes may deviate from optimal results in certain practical mMIMO configurations, necessitating the adoption of instantaneous CSI-based schemes. In addition, they may be limited in practical implementation where instantaneous CSI-based resource allocation and management schemes are widely adopted.
\IEEEPARstart{T}{his} lecture provides a comprehensive comparison between the statistical CSI-based power allocation and instantaneous CSI-based power allocation designs for mMIMO systems from \textit{performance, complexity, and practical implementation aspects}.
%
%
%
%
%
\section*{Relevance}
\label{sec:length}
Power allocation has long been regarded as an essential part of 
conventional wireless cellular networks. To date, extensive efforts have been undertaken to devise effective power control algorithms aimed at optimizing system performance across various objectives as well as providing uniformly good performance to the users. Nevertheless, due to the impact of  inter-user interference, which occurs when multiple users transmit simultaneously over shared communication resources, the majority of the proposed algorithms either involve advanced optimization techniques or rely  on some suboptimal algorithms which simplifies the computation. In addition, it is important to point out that most studies tend to investigate power allocation schemes concerning assumption of knowing instantaneous CSI of all links in the network. By adapting the power allocation coefficients on varying small-scale fading,  the \emph{instantaneous CSI-based power control} schemes provide significant improvements when there are large variations in channel quality over the subcarriers. However,  the required channel acquisition and computational overhead is prohibitive which scale with number of antennas, subcarriers, and users.

Currently, mMIMO technology has become a key factor in improving both spectral and energy efficiency in wireless networks. This improvement is achieved by employing a large number of antennas at each BS relative to the number of served users~\cite{marzetta2016fundamentals,Zhang:2014:JSTSP, Mohammadali:TCOM:2024}. This mature technology is now actively being integrated into 5G standards, marking a remarkable advancement in wireless communication systems. However, power control in  mMIMO systems is challenging and mainly nontrivial extensions of conventional wireless networks. In particular, utilizing the same power control principles used in conventional wireless networks would make optimal power control in  mMIMO systems even more challenging when  dealing with a large number of antennas. On the other hand, owing to the eminent advantage of channel hardening, power allocation can be significantly simplified  in mMIMO systems. The channel hardening phenomenon means that 
the variations of  channel gains  to the users—specifically, the squared norms of the channel vectors—are almost negligible over the frequency domain, and hence, the channel gains are almost deterministic~\cite{Björnson:BOOK:2017}. Therefore, the system performance is almost independent of the small-scale fading and mainly depends on the first and second order statistics (large-scale fading) of the mMIMO channels, also known as statistical CSI.  Accordingly,  by exploiting the channel hardening phenomenon, the whole spectrum can be simultaneously allocated to each user, and the statistical CSI-based power control can be performed jointly for all subcarriers relying only on the  large-scale fading time scale.  Interestingly, exploiting the unique hardening characteristic allows for obtaining closed-form expressions for the achievable rate of mMIMO systems in scenarios involving common fading models such as Rayleigh fading, using certain linear processing schemes, which makes it even easier to formulate and solve  power allocation problems.
In summary, the statistical CSI-based design can make advanced power allocation algorithms practically implementable.

While it is an important conclusion for power allocation design, in practice for some propagation environments, such as  keyhole channels, channel hardening breaks down~\cite{Matthaiou:SPAWC:2019}, hence  using the mean of the effective gain as substitute for its true value for power allocation  may result in poor performance. Furthermore,  in scenario associated with small to moderate numbers
of antennas, the level of channel hardening is lesser and the channel gains may still deviate significantly from their corresponding means. In turn, this reveals that  using instantaneous CSI-based power allocation  in mMIMO systems is inevitable for some  network setups. 
Therefore, a comparison between statistical CSI-based power allocation and instantaneous CSI-based power allocation designs for mMIMO systems across various aspects is both necessary and highly insightful for industry  and academia. While the theoretical foundations and distinct features of power control in mMIMO systems are well established in the literature~\cite{marzetta2016fundamentals,Björnson:BOOK:2017}, there is no single, comprehensive reference that thoroughly compares statistical CSI-based power allocation and instantaneous CSI-based power allocation designs in terms of \emph{performance}, \emph{complexity}, and particularly \emph{practical implementation} aspects. This lecture  aims to fill this gap by offering a detailed, side-by-side comparison that is not found in existing references, providing new insights into the practical challenges and trade-offs of each approach.

\emph{Notation}: In what follows, we use bold upper case letters to denote matrices, and lower case letters to denote vectors.  The superscript $(\cdot)^\dag$ stands for the conjugate-transpose (Hermitian), while $\mathbb{E}\{\cdot\}$ and  $\mathbb{C}^{L\times N}$ 
 denote the statistical expectation and an $L\times N$ matrix,  respectively. A circular symmetric complex Gaussian random variable (RV) having a variance of $\sigma^2$ is denoted by $\mathcal{CN}(0,\sigma^2)$.  
%
\section*{Prerequisites}
\label{secpr}
\par The prerequisites for understanding this
lecture note material are basic knowledge of probability theory, linear algebra,  convex optimization, communication theory, mMIMO, and signals and systems.

\section*{Problem Statement and Solutions}
We begin by briefly describing the transmission protocol in mMIMO systems and then formulating two classic power control problems with different information sharing requirements (i.e., statistical CSI or instantaneous CSI).  The problem at hand is to thoroughly compare the complexity and performance  of statistical CSI-based power allocation and instantaneous CSI-based  designs for different propagation environments. We will address the following question: Given a system requirement and propagation environment, which power control design offers the best complexity-performance trade-off? In addition, practical aspects are taken into consideration when we make the comparison.

Consider a single-cell mMIMO system, in
which one  BS  serves a set  of users, $\mathcal{K}\triangleq \{1,\dots,K\}$, located in the cell. The BS is equipped with $M$ antennas, $M>K$, while each user has a single antenna. 
The channel vector between the BS and the $k$-th user is denoted by  $\qg_k\in\mathbb{C}^{\Ntx \times 1}$,  which is modeled as 
\begin{align}\label{eq:gk}
    \qg_k=\sqrt{\beta_k}\qh_k,
\end{align}
where $\qh_k$ shows small-scale fading, which is caused by subtle changes in the propagation environment, such as the movement of the transmitter, receiver, or surrounding objects. The elements of $\qh_k$   are uncorrelated, zero-mean, and unit-variance RVs which are not necessarily Gaussian distributed\footnote{This general assumption allows for a broader class of fading models, such as Rayleigh and keyhole channels, which may exhibit different statistical properties.}. Moreover, for any $k \neq k'$,  $\qh_k$ and $\qh_{k'}$ are independent.
In addition, $\beta_k$ is the large-scale fading coefficient, which includes
distance-dependent path-loss, shadowing, and antenna gains~\cite{Björnson:BOOK:2017}.
Small-scale fading is assumed to remain static within each coherence interval and vary independently between coherence intervals. In contrast, large-scale fading changes more slowly and remains constant over several coherence intervals. The coherence interval, denoted as $\tau_c$, is the product of the coherence time (the duration over which the channel response remains approximately constant) and the frequency coherence interval (the frequency range where the channel response is approximately constant).

We assume that the users are served simultaneously within the same time-frequency resource. The system operates using a time division duplex (TDD) protocol, which means that the communication between the BS and the users is divided into separate time slots for the uplink (i.e., transmission  from users to the BS) and the downlink transmission (i.e., transmission from the BS to the users)~\cite{marzetta2016fundamentals}.
In particular, each coherence interval  is divided into an uplink training phase,  a downlink payload data transmission phase, and an uplink payload data transmission phase. Here, we focus on the downlink transmission, and hence, the uplink payload transmission phase is ignored.


\subsection*{Uplink Training  and Downlink Transmission Phases}
In the uplink training phase, all users send  pilot sequences to the BS and then BS estimates their corresponding channels. By relying on uplink/downlink channel reciprocity,  i.e., assuming the same channel response in both directions, the downlink CSI can be estimated through uplink training. 
By following~\cite{Hien:cellfree}, for the linear minimum-mean-square-error (MMSE) estimation technique and the assumption of orthogonal pilot sequences, the   estimate of $\qg_k$  can be obtained as
\begin{align}
\ghat_k=\frac{\tau_\mathrm{u}\rho_\mathrm{u}\beta_k}{\tau_\mathrm{u}\eta_\mathrm{u}\beta_k+1}\qg_k+\frac{\sqrt{\tau_\mathrm{u}\rho_\mathrm{u}}\beta_k}{\tau_\mathrm{u}\eta_\mathrm{u}\beta_k+1} \qn_k,
\end{align}
where $\qn_k \sim \mathcal{CN} (\bold{0}, \mathbf{I}_\Ntx$) is independent of $\qg_k$. Also, $\rho_\mathrm{u}$ and $\tau_\mathrm{u} \geq K$ are the normalized transmit power of each pilot symbol and the length  of pilot sequences, respectively.

In the downlink transmission phase, the BS transmits a separate signal to each user $k \in \mathcal{K}$ using linear precoding from an array of $M$ antennas. Precoding ensures that each data signal is transmitted from all antennas, but with varying amplitudes and phases to spatially direct the signal toward the intended user.
Denote by $s_{k}$  the symbol intended for  user $k$. We assume that $s_{k}$ is a RV with zero mean and unit variance.
The transmitted signal from the BS to the users, $\qx \in \mathbb{C}^{M \times 1}$, is generated by first scaling each symbol  $s_{k}$ with the power control coefficient $\eta_{k} \geq 0, \forall k$, and then multiplying them with the precoding vector $\qw_k$, where $\|\qw_k\|^2=1$, as
\begin{align}~\label{eq:x_l} 
		\qx=\sqrt{\rod} \sum_{k=1}^{K} \sqrt {\eta _{k}} \qw_k s_k,
\end{align}
where $\rod$ is the maximum normalized transmit power at the BS, i.e., $\Ex\{\|\qx\|^{2}\} \leq \rod$ which is equivalent to the power constraint
$\sum_{k\in\mathcal{K}} \eta_{k} \leq 1$.
The  signal received at the $k$-th user is  given by
	\begin{align}~\label{eq:z_k}   
		{z}_{k}=\underbrace{\sqrt{\rod}\sqrt {\eta _{k}} \qg^{\dag}_{k}\qw_{k} s_{k}}_{\text{Desired signal}}+\underbrace{\sqrt{\rod}\sum_{t=1, t\neq k}^{K} \sqrt {\eta _{t}} \qg^{\dag}_{k}\qw_{t} s_{t}}_{\text{Inter-user interference}}+{n}_{k},
	\end{align}
where $n_{k} \sim \mathcal{CN}(0,1)$.
We consider two common linear precoding schemes, namely maximum ratio (MR)  and zero forcing (ZF). For the MR scheme, the precoding vector $\qw_k$, $\forall k$, is designed as
\begin{align}\label{eq:mrt}
   \qw_k= \frac{\ghat_k^\dag}{\Vert\ghat_k^\dag\Vert}.
\end{align}
This precoding maximizes the desired signal by weighting each user's signal according to its channel's characteristics. For  ZF scheme, the precoding vector is designed as
\begin{align}\label{eq:zf}
   \qw_k= \frac{1}{\Big\Vert\big[\hat{\qG}(\hat{\qG}^\dag\hat{\qG})^{-1}\big]_k\Big\Vert}\Big[\hat{\qG}(\hat{\qG}^\dag\hat{\qG})^{-1}\Big]_k,
\end{align}
where $\hat{\qG}$ is the matrix of channel estimates, $\hat{\qG}=[\hat{\qg}_1,\cdots, \hat{\qg}_K]$. This precoding vector aims to eliminate inter-user interference by adjusting the precoding weights to nullify the signals from other users. Accordingly, for the use of  only statistical channel information,  an achievable rate of user $k$,  using a bounding technique~\cite{Björnson:BOOK:2017}, known as the hardening bound,  can be written as
 \begin{align}\label{eq:SE}
 \text {R}_k^\Stat=\log_2 (1+\SINRk^\Stat),
 \end{align}
 where
\begin{align}
\label{eq:SINRk}
\SINRk^\Stat= &\frac{\rod\eta_k a_k^\Stat}
{\rod\sum\limits_{t=1}^{K} \eta_tb_{tk}^\Stat+1}, 
\end{align}
with $a_k^\Stat =\big| \Ex\big\{\qg_{k}^{\dag} \qw_{k}\big\}\big|^{2}$, and
\begin{align}~\label{eq:Ex} 
b_{tk}^\Stat=
\begin{cases} \Ex\big\{\big| \qg_{k}^{\dag} \qw_{t}\big|^{2}\big\}, & t \neq{k}, \\ \Ex\big\{\big| \qg_{k}^{\dag} \qw_{k}\big|^{2}\big\}-\big|\Ex\big\{ \qg_{k}^{\dag} \qw_{k}\big\}\big|^{2}, & t ={k}. \end{cases}
\end{align}
  
The  term $b_{kk}^\Stat$, i.e., $\rod  \eta_k\Ex\big\{\big| \qg_{k}^{\dag} \qw_{k}\big|^{2}\big\}-\rod\eta_k\big|
\Ex\big\{\qg_{k}^{\dag} \qw_{k}\big\}\big|^{2}$ in the denominator of $\SINRk^\Stat$ is called the beamforming gain uncertainty~\cite{marzetta2016fundamentals}. This term arises from the user's lack of instantaneous channel gain knowledge, which reduces the achievable rate. However, it approaches zero with perfect CSI, where the instantaneous effective channel gain is fully known.

On the other hand, with perfect CSI,    achievable rate for  user $k$ is given by
\begin{align}\label{eq:perfect:SE}
\text {R}_k^\Ins=\Ex\{\log_2 (1+\SINRk^\Ins)\},
\end{align}
 where
\begin{align}
~\label{eq:Perfect:SINRk}
\SINRk^\Ins= \frac{\rod\eta_k a_k^\Ins}
{\rod\sum\limits_{t=1}^{K} \eta_tb_{tk}^\Ins+1}, 
\end{align}
with  $a_k^\Ins=\big|\qg_{k}^{\dag} \qw_{k}\big|^{2}$. Moreover, $b_{tk}^\Ins$ is equal to $\big| \qg_{k}^{\dag} \qw_{t}\big|^{2}$ for $t \neq{k}$, but otherwise is $0$. 

\subsection*{Importance of Channel Hardening}
One of the beneficial properties of mMIMO is channel hardening, which, in turn, makes statistical CSI-based power allocation deployment feasible. One motivation of this subsection is that mMIMO channels may not always harden, and hence, statistical CSI-based power allocation may not always be the best choice to achieve the highest performance.

\subsubsection*{Channel Hardening Definition and Advantages}
We now present the concept of channel hardening, wherein the inherent randomness of a fading channel is smoothed out through averaging, i.e., the squared norm of the combined channel vectors from the BS to the users remains relatively stable, even when the small-scale fading channels undergo random variations. From mathematical point of view, a propagation environment offers channel hardening if for any channel   $\qg_k$, we have~\cite{Björnson:BOOK:2017}
\begin{align}
    \frac{\Vert\qg_k\Vert^2}{\Ex\{\Vert\qg_k\Vert^2\}} \xrightarrow[{M \to \infty }]{a.s.} 1,
\end{align}
where $\xrightarrow[{M \to \infty }]{a.s.}$ denotes  almost sure convergence, as $M$ goes to infinity. When channel hardening holds for a particular propagation environment, the BS can replace the instantaneous channel gain by its mean value  for both signal detection and power allocation. Therefore, the power allocation can be greatly simplified, i.e., it can be implemented over the large-scale fading time  scale. \emph{Additionally,  downlink channel estimation phase for mMIMO systems relying on  channel hardening along with TDD operation is unnecessary, which  diminishes the channel estimation overhead. This is because each user can consider the average effective channel gain as the accurate one, allowing for the detection of the intended signal without the need for complicated channel estimation processes.}

On the other hand the authors in~\cite{Hien:TWC:2017,Matthaiou:SPAWC:2019} show that there are a number of scenarios under which the channel hardening property breaks down. Let us consider the following criterion introduced in~\cite{Hien:TWC:2017}  to assess whether channel hardening occurs or not:
\begin{align}
    \text{HC}_k=\frac{\text{Var}\{\Vert\qg_k\Vert^2\}}{\left(\Ex\{\Vert\qg_k\Vert^2\}\right)^2}. 
\end{align}
We can say that a  propagation environment offers channel  hardening if 
\begin{align} \label{eq:CH_mt}
\text{HC}_k \xrightarrow[]{ } 0 \quad \text{as} \quad {M \to \infty }
.\end{align}

\emph{1) Rayleigh Fading Channels}: When the entries of  $\qh_k$ in~\eqref{eq:gk} are i.i.d. $\mathcal{CN}(0,1)$ RVs, the channel model is referred to as i.i.d. Rayleigh fading. This terminology is used because the elements of $\qh_k$ are independent, with their magnitudes following a Rayleigh distribution. Rayleigh fading is a tractable model occurred in rich scattering conditions\footnote{Under rich (isotropic) scattering conditions, the signal experiences a high degree of multipath propagation, and the multipath components (different signal paths resulting from reflections, diffractions, or scattering) are uniformly distributed in all directions.}, where the BS antenna array is surrounded by many scattering objects relative to the number of antennas. Scattering occurs when the transmitted signal bounces off objects such as buildings or trees, creating multiple signal paths.
For independent Rayleigh fading channels, we have 
\vspace{-0.5em}
\begin{align}
    \text{HC}_k=\frac{1 } {M}  \xrightarrow[]{ } 0 \quad \text{as} \quad {M \to \infty }.
\end{align}
Thus, we have channel hardening. 

\emph{2)  Keyhole Channels:} The keyhole channel, or double scattering channel, refers to propagation environments where there is significant scattering around both the transmitter and receiver along with  a low-rank connection between the two scattering environments.  A `low-rank connection' between the two environments means that only a few dominant signal paths effectively contribute to the communication, which limits the system's ability to exploit the full diversity of the environment~\cite{Gesbert:TWC:2002}.
Two examples of communication scenarios with  keyhole effects are: 1) when radio signals go through  tunnels and corridors, and 2) when there exists a large separation between the transmitter and receiver without
line-of-sight (LoS) in a rich scattering environment. Let $o_k$ denote the number of effective keyholes for user $k$. Then, the keyhole channel can be modelled as
\vspace{-0.5em}
\begin{equation}
{\mathbf {g}}_{k} = \sqrt {\beta _{k}}\sum _{j=1}^{o_{k}} c_{j}^{(k)} p_{j}^{(k)} \qq_{j}^{(k)}, 
\end{equation}
where $\qq_{j}^{(k)}$ models an $M \times 1$ random channel vector between the $j$-th keyhole associated with the $k$-th user and has i.i.d. $~\mathcal{CN}(0,1)$ elements. Moreover, $p^{(k)}_j\sim \mathcal{CN}(0,1)$ shows the  channel from  $k$-th user to  $j$-th keyhole,  while $c_{j}^{(k)}$ is a deterministic value which denotes a complex gain of the $j$-th keyhole associated with the $k$-th user. We can conclude the following results for keyhole channels:
\begin{itemize}
\item Relying on the practical assumption that different users have different set of keyholes, we have~\cite{Hien:TWC:2017}
\vspace{-0.5em}
\begin{align}\label{keyhol:ch}
 \text{HC}_k=&\left ({1+\frac {1}{M}}\right ) \sum _{i=1}^{o_{k}} \left |{ c_{i}^{(k)}}\right |^{4} + \frac {1}{M},\notag \\\to&\sum _{i=1}^{o_{k}} \left |{c_{i}^{(k)}}\right |^{4} \neq 0, \quad M\to \infty,
 \end{align}
which shows that channels do not harden for keyhole environments.
\item It has also been shown that when $o_k=1$, i.e., we have single-keyhole channel,    the channel gain fluctuates the most and hence  single-keyhole channel can be regarded as the worst case. 
\item When $o_k=\infty$, we have Rayleigh fading.
\end{itemize}

\begin{tcolorbox}[
    title=\textbf{The First Lesson Learned:}, 
    colback=white, 
    colframe=black,
    colbacktitle=white,
    coltitle=black,
    coltext=black,
    boxrule=0.3mm
]
The channel-hardening property makes  instantaneous channel vector gain close to its mean value, and hence, power control can be designed by relying only on knowledge of that mean (statistical CSI). The criterion introduced in~\eqref{eq:CH_mt} can be effectively used to determine if channel hardening holds for a particular propagation environment. When $M$ is small the level of channel hardening becomes low. Also, in some propagation environments, such as keyhole channels, channel hardening does not hold.
  Therefore, the instantaneous channel vector gain is not close to its average.
  \end{tcolorbox}

\subsection*{ Power Control Problem Formulations and Solutions}

Now, we examine two power allocation problems for both statistical and instantaneous CSI-based designs, each subject to maximum transmit power constraints: 1) sum-rate maximization, where our goal is to maximize the total data rates of all users; and 2) max-min fairness, where we aim to improve the performance of the user with the lowest data rate. The solution to the max-min fairness problem ensures that all users receive a similar data rate performance.

\begin{enumerate}
    \item \textbf{Sum-Rate Maximization:}
Here, the power optimization problem can be formulated as
    \begin{subequations}\label{P:SE3}
	\begin{align}
		\underset{\boldsymbol{\eta}}{\mathrm{max}}\,\, &\hspace{1em}
		\underset{k\in\mathcal{K}} \sum \log_2(1+\SINRk^\sti(\boldsymbol{\eta}))
		\\
		\mathrm{s.t.} \,\,
		& \hspace{1em} \sum_{k\in\mathcal{K}} \eta_{k} \leq 1,\label{opt2:cons1}\\
  &\hspace{1em}\eta_{k}\geq 0, ~k\in\mathcal{K}, \label{opt2:cons2}
   \end{align}
\end{subequations}
where $\sti \in \{\Stat, \Ins\}$ and $\boldsymbol{\eta}\triangleq\{\eta_k, k=1,\cdots,K\}$. 
  This problem is generally  hard to solve. To   this end,    we use  Theorem 3 from  \cite{shen:TSP:2018} to transform     problem~\eqref{P:SE3} into a sequence of convex problems that update the power variables iteratively. In particular,   we can reformulate \eqref{P:SE3} as
    \begin{subequations}\label{P:SE4}
	\begin{align}
		\underset{\boldsymbol{\eta},\qy}{\mathrm{max}}\,\, &\hspace{1em}
		 f (\boldsymbol{\eta},\qy) 
		\\
		\mathrm{s.t.} \,\,
		& \hspace{1em} \sum_{k\in\mathcal{K}} \eta_{k} \leq 1,\label{opt2:cons1}\\
  &\hspace{1em}\eta_{k}\geq 0, ~k\in\mathcal{K}, \label{opt2:cons2}
   \end{align}
\end{subequations}
where $\qy\triangleq\{y_k, k=1,\cdots,K\}$ is a set of auxiliary variables with $y_k$ denoting an auxiliary variable associated with   $\SINRk^\sti$ and
\begin{align}\label{eq:f}
f (\boldsymbol{\eta},\qy)  = \sum _{k \in \mathcal{K}}  \log_2 \Bigg (1+2y_k\sqrt{ \rod\eta_{k} a_k^\sti}
 -y_k^2\Bigg (\sum _{t  \in \mathcal{K}}  \rod\eta_{t} b_{tk}^\sti +1 \Bigg)\Bigg). 
 \end{align}
We resort to \textbf{Algorithm~\ref{Alg:iterative}} to maximize $f (\boldsymbol{\eta},\qy)$ over $\boldsymbol{\eta}$ and $\qy$
in an iterative manner. More specifically, for the given $\eta_{k}$, the optimal $y_k^\star$ can be obtained from $\partial f/\partial y_k=0$,  as follows
\begin{equation}\label{eq:Ystar}
y ^\star _k = \frac{\sqrt{a_k^\sti\rod \eta_{k}}}{\sum _{t \in \mathcal{K}} b_{tk}^\sti \rod\eta_{t}+1}.
\end{equation}
In addition, for the fixed $\qy^\star$, $f (\boldsymbol{\eta},\qy^\star)$ is concave w.r.t. $\boldsymbol{\eta}$. Consequently, we can find the optimal $\boldsymbol{\eta}$ by solving the following convex optimization problem
    \begin{subequations}\label{P:SE4}
	\begin{align}
		\underset{\boldsymbol{\eta}}{\mathrm{max}}\,\, &\hspace{1em} f(\boldsymbol{\eta},\qy^\star) 
		\\
		\mathrm{s.t.} \,\,
		& \hspace{1em} \sum_{k\in\mathcal{K}} \eta_{k} \leq 1,\label{opt2:cons1}\\
  &\hspace{1em}\eta_{k}\geq 0, ~k\in\mathcal{K}. \label{opt2:cons2}
   \end{align}
\end{subequations}
The convergence of Algorithm~\ref{Alg:iterative} to a stationary point can be proved by using Theorem 4 in~\cite{shen:TSP:2018}.

\begin{algorithm}[t]
\caption{Iterative Approach for Solving Problem~\eqref{P:SE3} }
\begin{algorithmic}[1]
\label{Alg:iterative}
    \STATE \textbf{Initialize}:  Set  iteration index $\ell = 0$. Choose the initial value for $\boldsymbol{\eta}$  from the feasible set,  calculate $\qy$ using~\eqref{eq:Ystar} and then initialize $A^{(0)}=f(\qy,\boldsymbol{\eta})$ using~\eqref{eq:f}. Choose the maximum number of iterations, $L$, and tolerable accuracy $\epsilon > 0$. 
\REPEAT
    \STATE Update  $\qy$ by~\eqref{eq:Ystar}. 
\STATE Update optimum power allocation coefficients, $\boldsymbol{\eta}^\star$, through solving  Problem~\eqref{P:SE4} over $\boldsymbol{\eta}$ for fixed $\qy$.
\STATE Set $\ell=\ell+1$, $A^{(\ell)}=f(\qy^\star,\boldsymbol{\eta}^\star)$, and update $\boldsymbol{\eta}=\boldsymbol{\eta}^\star$.  
\UNTIL $|\frac{A^{(\ell)}-A^{(\ell-1)}}{A^{(\ell-1)}}| \leq \epsilon$ or $\ell = L$.
\end{algorithmic}
\end{algorithm}

\item \textbf{Max-Min Fairness Optimization}: Max-min fairness power optimization aims at maximizing the worst user's achievable rate, and presenting uniform service
throughout the network. Here, the power optimization problem can be formulated as
    \begin{subequations}\label{P:SE1}
	\begin{align}
		\underset{\boldsymbol{\eta}}{\mathrm{max}}\,\, &\hspace{1em}
		\underset{k\in\mathcal{K}} \min \,\, \log_2(1+\SINRk^\sti(\boldsymbol{\eta})) 
		\\
		\mathrm{s.t.} \,\,
		& \hspace{1em} \sum_{k\in\mathcal{K}} \eta_{k} \leq 1,\\
  &\hspace{1em}\eta_{k}\geq 0, ~k\in\mathcal{K}. 
   \end{align}
\end{subequations}
 By introducing the slack variable $\zeta$, we  reformulate~\eqref{P:SE1} as
    \begin{subequations}\label{P:SE2}
	\begin{align}
		\underset{\boldsymbol{\eta}}{\mathrm{max}}\,\, &\hspace{1em}
		 \zeta\\
  		\mathrm{s.t.} \,\,
		& \hspace{0.3em} \hspace{1em} \rod \sum\limits_{t \in \mathcal{K}} \eta_tb_{tk}^\sti+1-\frac{1}
{\zeta}\rod\eta_k a_k^\sti\leq 0, \label{opt1:cons1b}\\
		& \hspace{1em} \sum_{k\in\mathcal{K}} \eta_{k} \leq 1,\label{opt1:cons1}\\
  &\hspace{1em}\eta_{k}\geq 0, ~k\in\mathcal{K}. \label{opt:cons2}
   \end{align}
\end{subequations}
Constraints \eqref{opt1:cons1b} (for a given  $\zeta$) and \eqref{opt1:cons1}    are linear constraints with respect to the power-control coefficients $\eta_k$. Hence,  problem~\eqref{P:SE2}  is a standard quasi-linear optimization problem which can be solved by conducting a bisection search across the fixed parameter $\zeta$ and, for each $\zeta$,   solving a linear feasibility problem~\cite{Boyd:convex}.  The bisection method 
 used to obtain the solution of~\eqref{P:SE2} is summarized in \textbf{Algorithm~\ref{Alg:PA}}.
\end{enumerate}

\begin{algorithm}[t]
\caption{Max-Min Fairness Power Control }
\label{Alg:PA}
\begin{algorithmic}[1]
\STATE Initialize line-search accuracy  $\epsilon$, $\zeta_{\min}$, and $\zeta_{\max}$, where $\zeta_{\min}$ and $\zeta_{\max}$ define a range of minimum and maximum values of the objective function in~\eqref{P:SE1}. 
\REPEAT
    \STATE Set $\zeta:=\frac{\zeta_{\min}+\zeta_{\max}}{2}$. Solve the following convex feasibility program
  	\begin{align}\label{eq:bisection}
     \begin{cases} \hspace {0.0 cm}
		&   \sum\limits_{t \in \mathcal{K}} \rod\eta_tb_{tk}^\sti+1-\frac{1}
{\zeta}\rod\eta_k a_k^\sti\leq 0, ~~ \forall k\in\mathcal{K}\\
		& \sum_{k\in\mathcal{K}} \eta_{k} \leq 1,\\
  &\eta_{k}\geq 0, ~\forall\in\mathcal{K}. 
 \end{cases}
   \end{align}
\STATE If problem~\eqref{eq:bisection} is feasible, then set $\zeta_{\min}:=\zeta$, else set $\zeta_{\max} :=\zeta$.
\UNTIL{ $\zeta_{\max}-\zeta_{\min}<\epsilon$ }
\end{algorithmic}
\end{algorithm}
\section*{Instantaneous versus Statistical Power Control Comparison}
To systematically compare instantaneous and statistical CSI-based power control designs, we categorized the comparison into the following three aspects:

\subsection*{ Comparison from System Performance Aspects}
 We  consider a single-cell mMIMO system with radius $D$, in
which a  BS  comprising $M$ antennas serves $K$ users. The BS is located at the centre and users are placed uniformly at random inside the cell. We choose a system bandwidth of $B = 20$ MHz and the maximum transmit power at the BS is  $1$ W, while the  normalized maximum transmit powers $\rho_d$ can be calculated upon dividing this power by the noise power of $\sigma_n^2=-92$ dBm~\cite{marzetta2016fundamentals}. The three-slope model determines the path loss as described in ~\cite{Hien:cellfree}, where the path loss exponent varies based on the distance between BS and $k$-th user, $d_k$. It is $3.5$ if $d_k>50$ m, $2$ if $10~\text{m}<d_k\leq 50~\text{m}$ , and $0$ if $d\leq 10$ m.
To focus only on the effect of instantaneous/statistical CSI-based power allocation design on the system performance, we consider the perfect CSI case, though  similar results can be obtained for the imperfect CSI case. In addition, to account for the loss due to uplink training phase, we  examine the downlink net throughput which is defined as
\begin{align} \label{eq:net-throughput}
    \mathcal{S}^\sti=B\varphi^\sti\text{R}^\sti,
\end{align}
in which the channel estimation overhead has been taken into account.
More specifically, the terms $\varphi^\Stat\triangleq1-\frac{\tau_\text {u}}{\tau_\text {c}}$ and $\varphi^\Ins\triangleq1-\frac{\tau_\text {u}+\tau_\text{d}}{\tau_\text {c}}$   illustrate  that, for each coherence
interval of length $\tau_\text {c}$ samples, in the mMIMO
systems relying on  statistical CSI for detecting the
transmitted signals at the users, we spend $\tau_\text {u}$ samples for the uplink training, while in the mMIMO
systems relying on instantaneous CSI, we spend $\tau_\text {u}+\tau_\text{d}$ samples for the uplink and downlink training. We choose
 $\tau_\text {u}=\tau_\text {d} = K$, and $\tau_c = 200$ samples, corresponding to a coherence
bandwidth of $200$ KHz and a coherence time of $1$ ms. 


\begin{figure*}
\centering 
\begin{minipage}{.45\textwidth} 
\centering 
\includegraphics[width=1\linewidth]{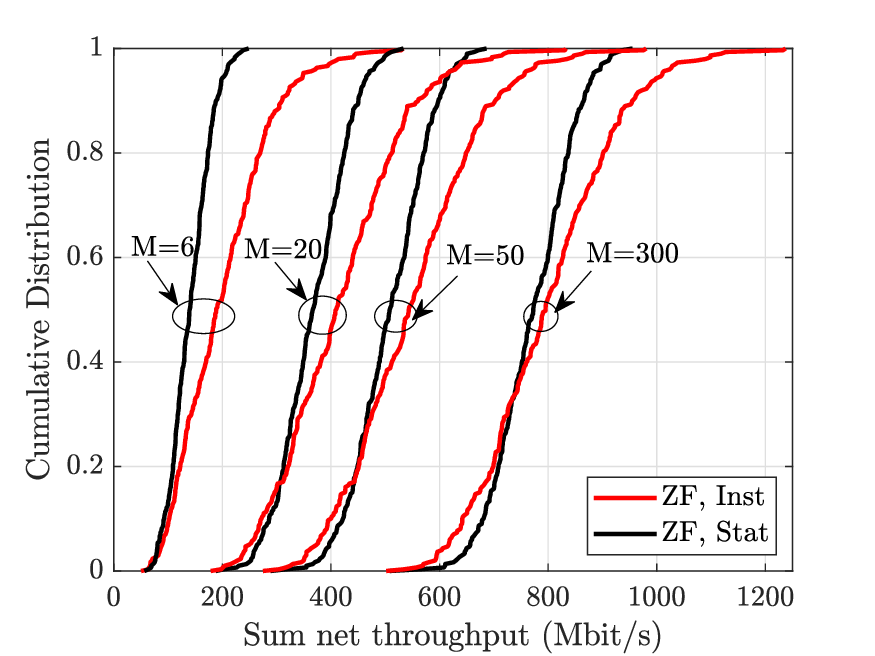}
\subcaption {ZF processing, Rayleigh fading channels.}
\label{fig1_a} 
\end{minipage}%
\begin{minipage}{.45\textwidth} 
\centering \includegraphics[width=1\linewidth]{ 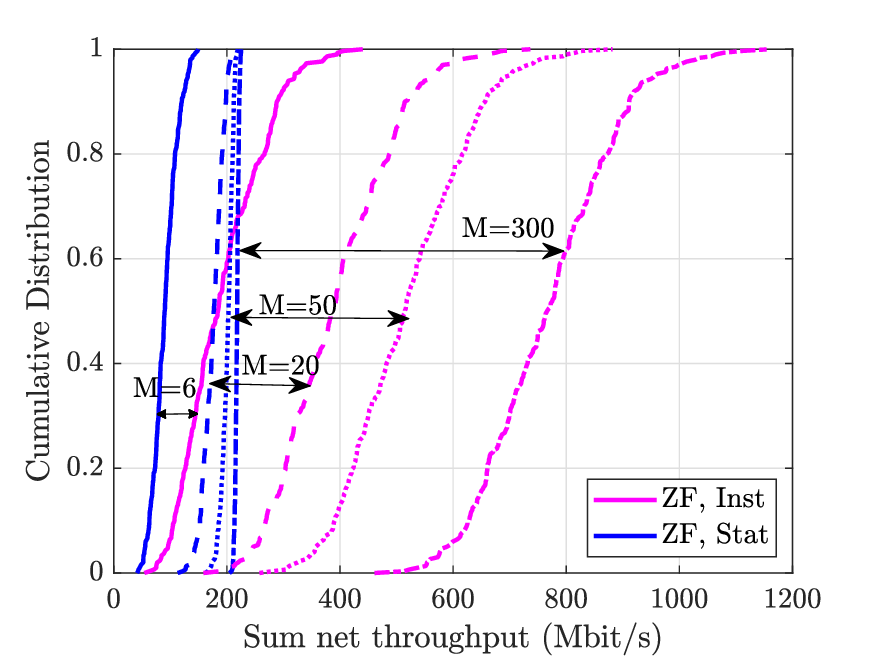}
\subcaption {ZF processing, single-keyhole channels.}
\label{fig1_b}
\end{minipage} 
\begin{minipage}{.45\textwidth} 
\includegraphics[width=1\linewidth]{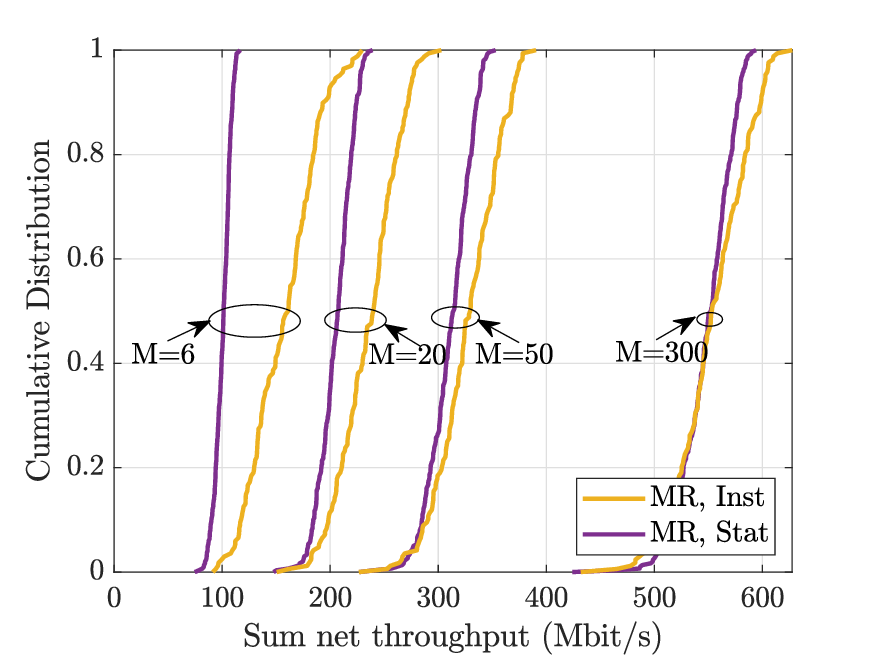}
\subcaption {MR processing, Rayleigh fading channels.}
\label{fig1_c} 
\end{minipage} 
\begin{minipage}{.45\textwidth} 
\centering \includegraphics[width=1\linewidth]{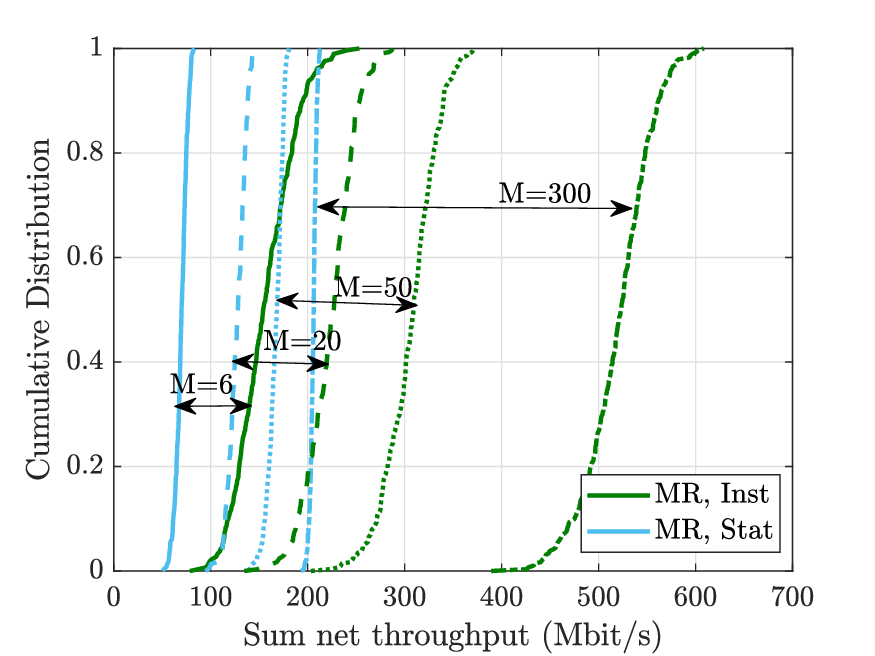}
\subcaption {MR processing, single-keyhole channels.}
\label{fig1_d} 
\end{minipage} 
\caption{The cumulative distribution of the sum downlink net throughput for ZF and MR processing with Algorithm~\ref{Alg:iterative}, where  $K=5$ and  $D=500$ m. 
}\label{fig1}
\hrulefill
\end{figure*}

In Fig.~\ref{fig1}, we compare the sum net throughput performance of  the statistical CSI-based and instantaneous CSI-based power allocation designs achieved by~\textbf{Algorithm~\ref{Alg:iterative}}  for  ZF and MR processings with different number of antennas under Rayleigh fading and single-keyhole channels. From this figure, we have the
following observations.

\begin{itemize}
\item In Rayleigh fading channels with a high number of antennas, for both ZF  and  MR processing,  the sum-net throughput performance gap  between statistical CSI-based and instantaneous CSI-based power allocation designs is very small, which means that using  the mean of the effective gain instead of the true channel gains for power allocation works very well and hence the instantaneous CSI-based design is not  necessary.

\item The disadvantage of statistical CSI-based power allocation in Rayleigh fading channels with a small  number of antennas  is clear. In particular, when $M=6$,  the switch from instantaneous CSI-based power allocation  to its statistical CSI-based counterpart  entails $35\%$ and $60\%$ reduction in sum-net throughput performance in terms of median  for ZF and MR processing, respectively. This behaviour follows from the fact that the level of channel hardening remarkably decreases for smaller $M$. 

\item Instantaneous CSI-based design suffers  from  higher channel estimation overhead as an extra $\tau_\text{d} \geq K$ samples need to be used for  downlink training. More specifically, for this simulation setup,  the pre-log factor in~\eqref{eq:net-throughput} is equal to $\varphi^\Stat=0.975$ for statistical CSI-based design, while it reduces to  $\varphi^\Ins=0.95$ for instantaneous CSI-based design. This leads to poor performance, especially
at low percentiles and large $M$, where instantaneous CSI-based design performs slightly worse than statistical CSI-based design. 
This highlights the disadvantage of instantaneous CSI-based design from the channel estimation overhead point of view, which becomes more pronounced  for the higher user loads.

\begin{figure}[!t]
			\centering 
			\includegraphics[width=0.5\textwidth]{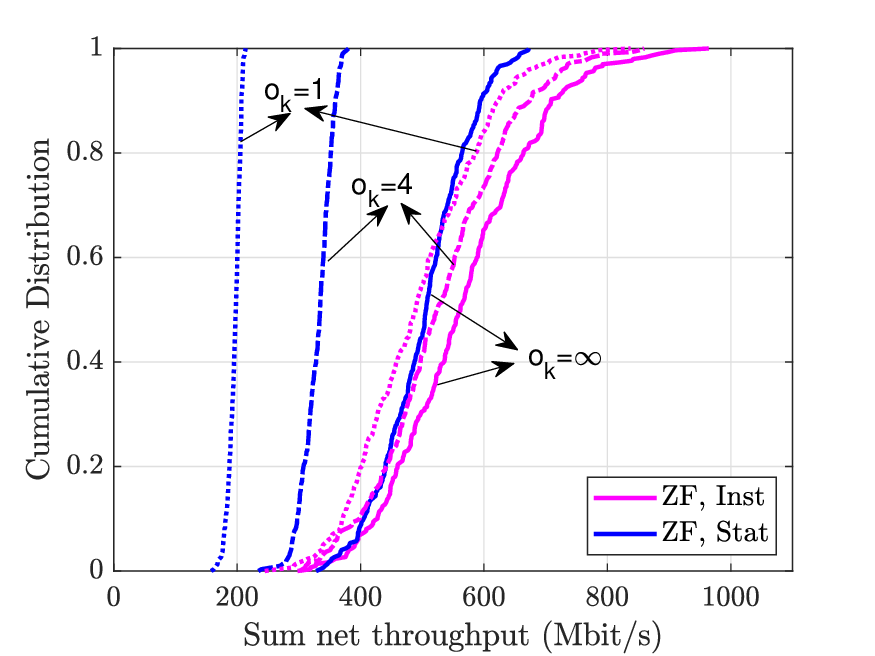}
			\caption{The cumulative distribution of the sum downlink net throughput for ZF processing with Algorithm~\ref{Alg:iterative} for different number of keyholes, $o_k$, where  $M=50$,  $K=5$ and $D=500$ m. }
			\label{fig1_2}
\end{figure}


\begin{figure*}
\centering 
\begin{minipage}{.45\textwidth} 
\centering 
\includegraphics[width=1\linewidth]{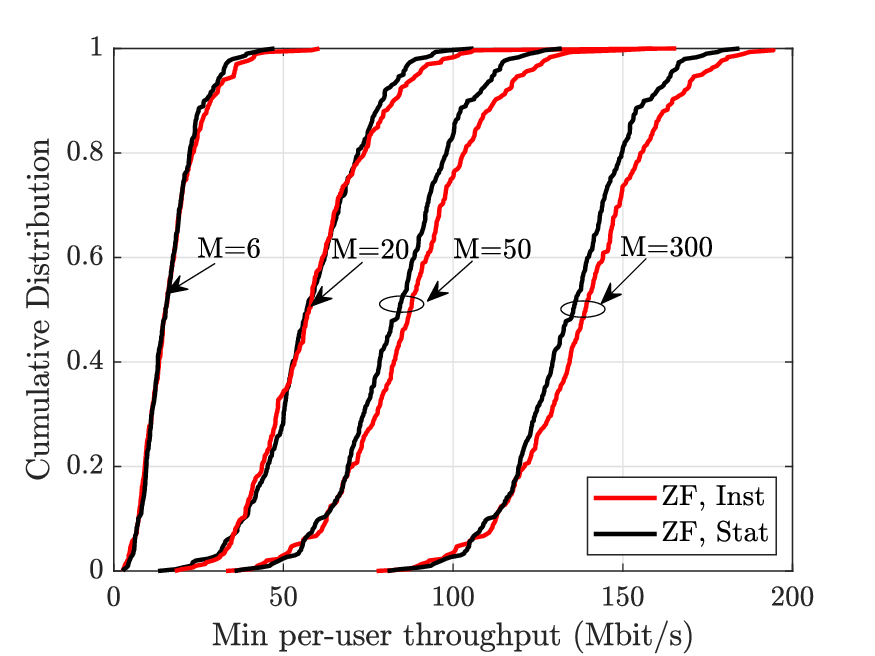}
\subcaption {ZF processing, Rayleigh fading channels.}
\label{fig2_a} 
\end{minipage}%
\begin{minipage}{.45\textwidth} 
\centering \includegraphics[width=0.95\linewidth]{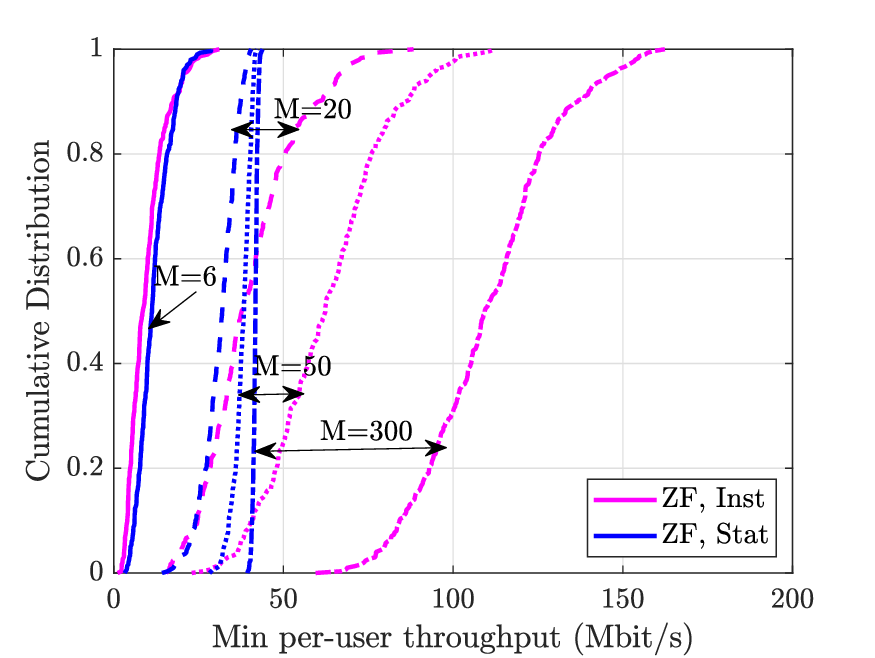}
\subcaption {ZF processing, single-keyhole channels.}
\label{fig2_b}
\end{minipage} 
\begin{minipage}{.45\textwidth} 
\includegraphics[width=1\linewidth]{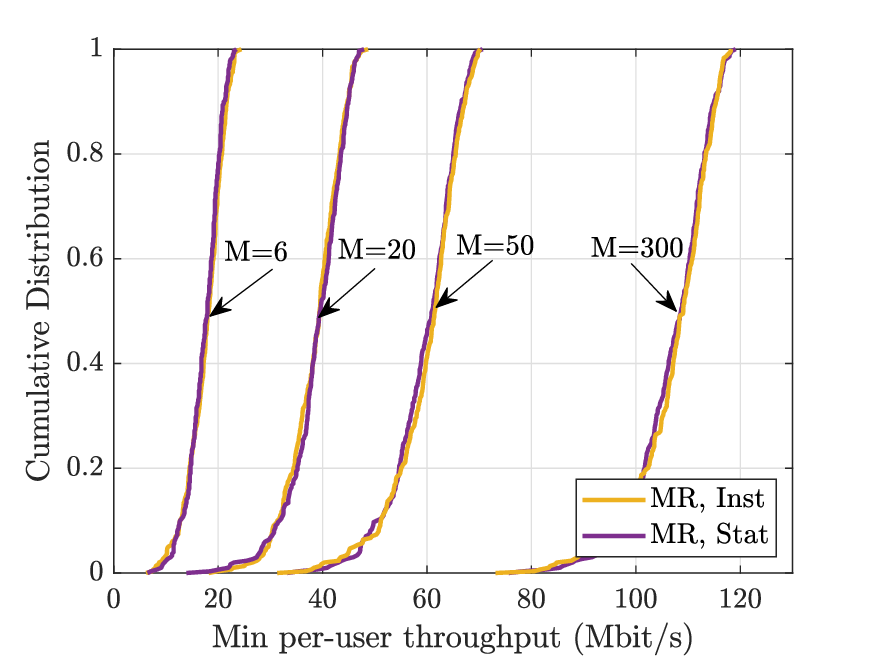}
\subcaption {MR processing, Rayleigh fading channels.}
\label{fig2_c} 
\end{minipage} 
\begin{minipage}{.45\textwidth} 
\centering \includegraphics[width=1\linewidth]{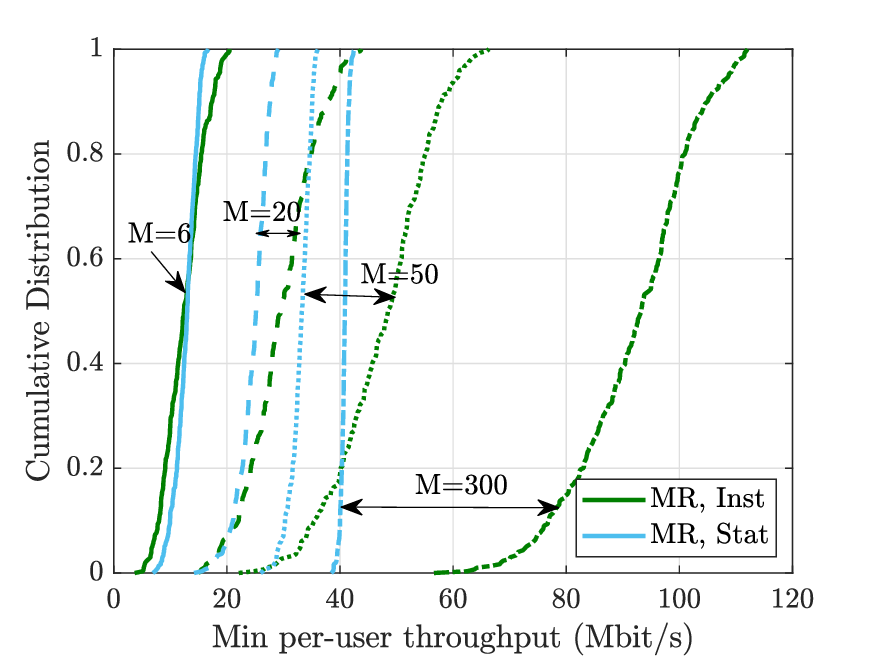}
\subcaption {MR processing, single-keyhole channels.}
\label{fig2_d} 
\end{minipage} 
\caption{The cumulative distribution of the minimum per-user downlink net throughput for ZF and MR processing with Algorithm~\ref{Alg:PA}, where  $K=5$ and $D=500$ m. 
}\label{fig2}
\hrulefill
\end{figure*}

\item  In keyhole channels, statistical  CSI-based power allocation significantly performs worse than instantaneous CSI-based scheme  even in the regime of high values for $M$. In fact, we can see that the relative performance gaps between  statistical  CSI-based and instantaneous CSI-based designs increase with $M$. The reason is that  in the keyhole propagation environment with a finite number of keyholes, $o_k$, the channels do not harden regardless of the value of $M$. Therefore, as observed from \eqref{eq:SINRk}, both the beamforming uncertainty and the desired signal components of $\SINRk^{\Stat}$  increase at the same rate as 
$M$ increases. This scaling effect limits the performance of statistical CSI-based designs for higher values of $M$.    
\end{itemize}

Next, we study the effects of the the number of keyholes $o_k$ on the performance of instantaneous and statistical CSI-based power allocation designs in Fig.~\ref{fig1_2}. It can be seen that  the performance gap between statistical CSI-based and instantaneous CSI-based power allocation designs reduces upon increasing the number of keyholes. More specifically, when we increase $o_k$ from $4$ to $\infty$ the performance gap between these two designs reduces from $56\%$ to $10\%$ in terms of median sum net throughput.

Here, we compare the minimum per-user net throughput performance of max-min fairness statistical CSI-based and instantaneous CSI-based power allocation designs achieved by~\textbf{Algorithm~\ref{Alg:PA}} under Rayleigh fading and single-keyhole channels for  ZF and MR processing with different number of antennas. In particular, Fig.~\ref{fig2} shows the cumulative distribution of the minimum per user downlink net throughput   for $D=500$ m. 

\begin{itemize}
\item For MR processing, in Rayleigh fading channels,  the performance gap  between statistical CSI-based and instantaneous CSI-based power allocation  is very small  regardless of the number of transmit antennas, which means that using  the mean of the effective gain instead of the true channel gains for max-min fairness power allocation in mMIMO systems relying on MR processing works very well and hence the instantaneous CSI-based design is not  necessary. In principle, this is reasonable because in the mMIMO system relying on MR processing, the inter-user interference is a more dominant factor for the performance of the worst user than the channel hardening level.
By contrast, in the mMIMO system relying on ZF processing, for perfect uplink estimation, the interference towards all the users can be completely cancelled out and hence when the short term average power constraint is applied the level of hardening  is predominant.

\item  In single-keyhole channels, max-min fairness statistical  CSI-based power allocation significantly performs worse than instantaneous CSI-based design  when number of antennas  $M$ is high. This is because, in single-keyhole channels, the channels do not harden and hence, both the beamforming uncertainty and the desired signal components of $\SINRk^{\Stat}$ increase at the same rate when $M$ grows large which in turns restricts the minimum per user throughput of statistical CSI-based designs.
However, the minimum per user net throughput performance of statistical CSI-based  designs is comparable to those of instantaneous CSI-based designs when $M$ is  small (e.g., $ M\leq 20$).
\end{itemize}

\begin{tcolorbox}[
    title=\textbf{The Second Lesson Learned:}, 
    colback=white, 
    colframe=black,
    colbacktitle=white,
    coltitle=black,
    coltext=black,
    boxrule=0.3mm
]
Instantaneous CSI-based design can be implemented in any propagation environment without relying on any particular channel statistics or the number of antennas. By way of contrast, when the sum net throughput is the objective of interest, statistical CSI-based power allocation cannot perform  well in non-hardening propagation environments such as keyhole channels or Rayleigh channels in scenarios associated with small to moderate number of antennas. On the other hand, with max-min fairness optimization, the performance of statistical CSI-based power allocation  is fairly close to instantaneous CSI-based schemes for a wide range of number of antennas. 
  \end{tcolorbox}

\subsection*{Comparison from Computational Complexity Aspects}
We now provide a brief comparison between the computational complexity requirements of instantaneous CSI-based design and those of statistical CSI-based design. 
\begin{itemize}
\item \emph{Overhead}:
In instantaneous CSI-based designs, the power allocation is recomputed on the small-scale fading time scale (instantaneous channels).  In contrast, for  statistical CSI-based design the  power allocation is recomputed on the large-scale fading time scale (statistical property of the channels). Note that small-scale fading coefficients change quickly over times and frequencies, while large-scale fading coefficients stay constant over frequencies and charge slowly with time compared to the small-scale fading (i.e., the large-scale fading changes can be at least  40 times slower according to~\cite{Ashikhmin:TIT:2018}). 

\item \emph{Centralized/Distributed Implementation}: Instantaneous CSI-based power allocation  requires a centralized implementation  at the central processing units (CPUs). Specifically, the BSs and users have to send all channel estimates to the CPUs which will cause very large overhead. Statistical CSI-based designs can be done distributed at the BSs and BSs need to know only the statistical property of the channels. 

\item \emph{CSI Knowledge at the Users}:  Instantaneous CSI-based designs require instantaneous CSI knowledge at the users, which causes a very large overhead in systems with many users. However, statistical CSI-based designs do not require instantaneous CSI knowledge at the users, making the design  scalable with respect to the number of users.
\end{itemize}


  \textbf{Practical Example}: Let us consider 5G  new radio (NR)  structure: bandwidth is $100$ MHz, subcarrier spacing is $15$ KHz, and frame length is $10$ ms which includes $10$ transmit time intervals (TTIs). Each frequency granularity (equivalently to the coherence bandwidth) includes $24$ subcarriers, i.e., $2$ resource blocks (RBs) (number of subcarriers per RB is $12$). Thus, frequency granularity = $24\times15 = 360$ KHz. For a bandwidth of $100$ MHz, we have $100\times10^3/360 = 277$ frequency granularity intervals. 
    Therefore, \emph{in mMIMO system relying on instantaneous CSI-based design, in total, the power allocations need to be computed $277\times10 = 2770$ times for each frame duration.} Note that even for one TTI, the power allocations need to be computed $277$ times (corresponding for each frequency granularity interval).   However, with the statistical CSI-based designs, we need to perform only $1$ time, and use the results for all subcarriers and several frame durations.   Note that if the system remains static  (the users are unchanged) for a long time, then the results for the statistical CSI-based designs can be used for about $10$ frames (because the large-scale fading coefficients can be stay constant for $100$ ms). With $10$ frames, the instantaneous channel-based designs have to perform the power allocations $27700$ times which is huge.


   \subsection*{Comparison from  Practical  Aspects}
   While statistical CSI-based power control algorithms rely on the  knowledge of large-scale fading coefficients (LSFCs) as deterministic variables, their practical implementation encounters  challenges.  In practice, it is not easy to  measure the LSFC from all  users, especially when they stay silent for a prolonged periods and move or the propagation conditions change~\cite{Fengler:2021:J_IT}. Nonetheless, LSFCs typically demonstrate slow variations in many real-world scenarios. Conversely, instantaneous CSI-based designs necessitate estimates of the the channels (i.e., combined channels of small-scale and large-scale fading) rather than LSFCs.

Moreover, in the current implementation of mMIMO techniques within practical 5G networks, the industry favors instantaneous CSI-based power allocation schemes. This preference arises from the fact that user scheduling is currently performed with each TTI. Consequently, the set of active users associated with a particular BS may vary from one TTI to another, especially when system dynamics change rapidly, to facilitate both load balancing and proportional fairness among users. Therefore, recalculating power allocation, whether based on statistical or instantaneous CSI, is necessary for each TTI. Nevertheless, if the set of active users remains the same as in the previous TTI, we should not recompute the power allocation. Therefore, in any case, it is still more beneficial to use the statistical CSI-based schemes. If the set of active users changes compared to the previous TTI, we can re-run the statistical CSI-based power allocation algorithm. Otherwise, we can use the results obtained in the previous TTI.

  Table~\ref{tabel: CSI_compare} boldly and explicitly summerizes the system performance, computational complexity, and practical aspects comparison of instantaneous and statistical CSI-based power allocation algorithms.

   

   \begin{center}
\begin{table*}
    \centering
\caption{Comparison  of instantaneous and statistical CSI-based power allocation schemes in terms of system performance, computational complexity, and practical implementation.}
    \label{tab:example}
    \begin{tabular}{| >{\centering\arraybackslash}m{0.5cm} || >{\centering\arraybackslash}m{7cm} || >{\centering\arraybackslash}m{7cm} |}
        \hline
        \textbf{ } & \textbf{Instantaneous CSI-based Schemes} & \textbf{Statistical CSI-based Schemes} \\
        \hline 
        \hline 
        \multirow{0 }{*}[0.2em]{\rotatebox[origin=c]{90}{System Performance}} & \multicolumn{2}{>{\centering\arraybackslash}m{14cm}|}{With sum-rate optimization:  

        1) In Rayleigh fading channels with a high number of antennas  the   performance gap between statistical CSI-based and instantaneous CSI-based power allocation designs is very small.
        
        2) Instantaneous CSI-based power allocation outperforms statistical CSI-based scheme in non-hardening propagation environments such as keyhole channels or Rayleigh channels in scenarios associated with small  number of antennas.} \\
        \cline{2-3}
         & \multicolumn{2}{>{\centering\arraybackslash}m{14cm}|}{With max-min fairness optimization: 
         
        1) The performance of statistical CSI-based power allocation is fairly close to instantaneous CSI-based schemes  in Rayleigh fading channels for a wide range of number of antennas.
        
       2) In single-keyhole channels,  statistical  CSI-based power allocation  performs worse than instantaneous CSI-based design, especially  when number of antennas  $M$ is high.
                } \\
        \hline
        \hline
        \multirow{0 }{*}[0.6em]{\rotatebox[origin=c]{90}{Complexity}} & The power allocation is recomputed on the small-scale fading time scale (instantaneous channels). & The power allocation is recomputed on the large-scale fading time
scale (statistical property of the channels). \\
        \cline{2-3}
         & Need to be done at the CPUs & Can be done distributed at the BSs. \\
        \cline{2-3}
         & Require instantaneous CSI knowledge at the users. & Do not require instantaneous CSI knowledge at the users. \\
        \hline
        \hline
        \multirow{0 }{*}[  -0.5em]{\rotatebox[origin=c]{90}{Practical Aspects}} & There is no need to estimate large-scale fading coefficients; only the channels (i.e., the combined channels of small-scale and large-scale fading) need to be estimated  & Necessitate estimates of the large-scale fading.  \\
        \cline{2-3}
         & The industry favors instantaneous CSI-based power allocation schemes due to the fact that user scheduling is currently done for each TTI. & Statistical CSI-based schemes can be more beneficial as they can be applied for each TTI or for several TTIs, depending on the application scenario. \\
        \cline{2-3}
         & Can be applied for different propagation environments. & Only perform well for the propagation environments with the hardening property (i.e., $\text {HC}_k \xrightarrow[]{ } 0 \quad$ \text{as} $\quad {M \to \infty}$). \\
        \hline
    \end{tabular}
     \label{tabel: CSI_compare}
\end{table*}
\end{center}
\section*{What We Have Learned}
This “Lecture Notes” article described and compared statistical CSI-based power allocation and instantaneous CSI-based power allocation designs for mMIMO systems. The main message is that statistical CSI-based designs provide better  trade-offs between performance, complexity, and signaling overhead than instantaneous CSI-based designs in hardening environments for both ZF and MR processing. Interestingly, even in Rayleigh fading channels with low level of channel hardening,   the performance of statistical CSI-based max-min fairness optimization relying on MR processing  is very close to its instantaneous CSI-based counterpart.
On the other hand, the disadvantage of statistical CSI-based design in non-hardening propagation environments is non-negligible for some scenarios. In particular, in mMIMO systems relying on either ZF or MR processing with the objective of sum-rate maximization,  the choice of instantaneous CSI-based design is crucial to achieve the potential gains of power control in either non-hardening scenarios or scenarios associated with small number of transmit antennas at BS.  Regarding practical aspects, since user scheduling is done for each TTI, instantaneous CSI-based schemes are preferable in practice. However, it is worth noting that statistical CSI-based schemes are still more beneficial because they can be applied for each TTI or for several TTIs, depending on the scenario.



\section*{Acknowledgment}
This work was supported by the U.K. Research and Innovation Future Leaders Fellowships under Grant MR/X010635/1, and a research grant from the Department for the Economy Northern Ireland under the US-Ireland R\&D Partnership Programme.

\bibliographystyle{IEEEtran}
\bibliography{IEEEabrv,references}

\section*{Biographies}
\label{sec:bio}

\begin{IEEEbiographynophoto}{Zahra Mobini}
(zahra.mobini@qub.ac.uk)  received the Ph.D.
degrees in electrical engineering from the  K. N. Toosi University of Technology, Tehran, Iran. From November 2010 to November 2011, she was a Visiting Researcher at the Australian National University, Australia. She is currently  a Post-Doctoral Research Fellow at Queen's University Belfast, U.K. Before joining QUB,  she was an Assistant and then Associate Professor with the Faculty of Engineering, Shahrekord University, Shahrekord, Iran (2015-2021). 
Her research interests include physical-layer security, massive  MIMO, cell-free massive  MIMO, full-duplex communications, and resource management and optimization. She is an IEEE member.
\end{IEEEbiographynophoto}
\vspace{11pt} 

\begin{IEEEbiographynophoto}{Hien Quoc Ngo}
(hien.ngo@qub.ac.uk, corresponding author) is currently a Reader with Queen's University Belfast, U.K, and a visiting professor at Kyung Hee University, South Korea. He has co-authored many research papers in wireless communications and co-authored the Cambridge University Press textbook "Fundamentals of Massive MIMO" (2016). He received the IEEE ComSoc Stephen O. Rice Prize in 2015, the IEEE ComSoc Leonard G. Abraham Prize in 2017, and the IEEE CTTC Early Achievement Award in 2023. He was awarded the UKRI Future Leaders Fellowship in 2019. He served/served as the Editor for the IEEE Transactions on Wireless Communications, the IEEE Transactions on Communications, the IEEE Wireless Communications Letters, the Digital Signal Processing, and Physical Communication (Elsevier). He is a Fellow of the IEEE.
\end{IEEEbiographynophoto}
\end{document}